# Plasma Astrophysics Problems in Star and Planet Formation

*White paper prepared for Astro2010 by Ellen Zweibel, Jeremy Goodman, Hantao Ji, and Alex Lazarian. Endorsed by the Center for Magnetic Self-Organization in Laboratory & Astrophysical Plasmas (www.cmso.info./), an NSF Physics Frontier Center in partnership with DoE, and GPAP, the*
*Topical Group in Plasma Astrophysics of the APS (www.aps.org/units/gpap/).*

The major questions relevant to star and planet formation are: *What controls the rate, efficiency, spatial clustering, multiplicity, and initial mass function of star formation, now and in the past? What are the major feedback mechanisms through which star formation affects its environment? What controls the formation and orbital parameters of planets, especially terrestrial planets?* These questions cannot be fully addressed without understanding key magnetohydrodynamics (MHD) and plasma physics processes. Although some of these basic problems have long been considered intractable, attacking them through a combination of laboratory experiment, theory, and numerical simulation is now feasible, and would be fruitful. Achieving a better understanding of these processes is critical to interpreting observations, and will form an important component of astrophysical models. These models in turn will serve as inputs to other areas of astrophysics, e.g. cosmology and galaxy formation.

Here are some examples of how MHD and plasma physics are embedded in the major questions posed above. Magnetic support in molecular clouds and cold cores, angular momentum transport by magnetic torques in protostellar disks, and the nature of magnetized turbulence in both of these environments (McKee et al. 1993, Mac Low & Klessen 2004, Elmegreen & Scalo 2004, McKee & Ostriker 2007) must be factors in *what controls the rate, efficiency, spatial clustering, multiplicity, and initial mass function of star formation, now and in the past*. The role of magnetic fields in launching and collimating outflows from young stellar objects and their circumstellar disks, and in transferring momentum and energy to the ambient medium (Krumholz et al. 2006, Banerjee et al. 2007, Matzner 2007), bears on *the major feedback mechanisms through which star formation affects its environment*. The transport of angular momentum in disks, the effect of hard radiation due to pre-main-sequence stellar magnetic activity, and the effect of magnetic fields on dust dynamics and dust-gas interaction are all important factors in *what controls the formation and orbital parameters of planets, especially terrestrial planets.* "Dead zones" in parts of protostellar disks where MHD turbulence is quenched, for example, may be necessary for dust to settle and agglomerate into planetesimals, or to slow orbital migration of planets after they form (Matsumura et al. 2007, Oishi et al. 2007). Interruption of the disk by the protostellar magnetosphere may account for the abundance of exoplanets in few-day orbits (Lin et al. 1996).

The basic processes which underlie the role of magnetic fields are the nature of magnetic turbulence, dynamo processes, magnetic reconnection, and angular momentum transport mediated by magnetic fields, all taking place in a weakly ionized, dusty medium. In protoplanetary disks, the electrical conductivity includes important Hall and Pedersen (ion-neutral drag) terms. Furthermore, the ionization itself is largely due to nonthermal processes involving magnetic fields, e.g. X-rays from protostellar coronae and cosmic rays accelerated in magnetized shocks.

Prospects for improved understanding of these issues in the coming decade derive from present trends in theory, computation, observations, and even laboratory experiments. The roles of microinstabilities and Hall terms in fast magnetic reconnection, for example, have been illuminated in recent years by dedicated experiments, simulations, and theoretical insights (Yamada et al. 2005)---but there is still a long way to go to reach astrophysically relevant regimes of dimensionless parameters. Turbulent magnetic dynamo action is being sought through simulations and experiments in resistive regimes relevant to planetary cores and protostellar disks, if not stellar interiors. Infrared and submillimeter molecular spectroscopy has begun to probe physical conditions in the planet-forming regions of protostellar disks; upcoming facilities such as ALMA and SOFIA promise to accelerate this progress and allow current models of disk turbulence, as well as earlier stages in star formation, to be tested and refined.

Addendum: the role of laboratory experiments

Observation, theory, and computation are standard tools of astronomy and astrophysics. The role of laboratory experiments outside atomic and molecular physics is perhaps less well known, so we briefly discuss one example with which we are familiar. This is the liquid metal MRI experiment at Princeton, which studies magnetorotational and other instabilites in the resistive regime appropriate to portions of protoplanetary disks.

A demonstration that a tabletop liquid gallium experiment could capture magnetorotational instability (MRI) in astrophysically interesting parameter regime was provided in Ji et al. (2001). Numerical simulations reported in Obabko et al. (2008) showed that the experiment as originally conceived would develop strong Ekman flows which would compete with the MRI,. This led to a modification of the experiment at the boundaries, which eliminated the problem.

The experiment was first run as a hydrodynamics experiment, with water instead of gallium. In this mode, it tested a prediction that unmagnetized Keplerian disks are unstable to finite amplitude disturbances which transport angular momentum, but obtained a null result (Ji et al. 2006).

The experiment is now running with gallium; an overview and update was presented to the astronomical community in Ji et al. (2008). Plasma MRI experiments, which would probe the instability at lower resistivity, possibly with the Hall effect and/or ion-neutral friction, thus capturing other aspects of protoplanetary disk physics, are in the planning stage.